%% file: sample-sigconf.tex
\documentclass[sigconf]{acmart}
\usepackage{soul}
\usepackage{xcolor}
\usepackage[most]{tcolorbox}
\usepackage{bm}
\usepackage{mdframed}
\usepackage{enumitem}
\usepackage{balance}
\usepackage{graphicx}
\usepackage{booktabs} 
\usepackage{multirow}
\usepackage{multicol}
\usepackage{colortbl}
\usepackage{makecell}
\usepackage{graphicx}
\usepackage{amsmath}
\usepackage{booktabs} 
\usepackage{multirow} 
\usepackage{array} 
\usepackage{subcaption} 
\usepackage{float} 
\usepackage{booktabs}
\AtBeginDocument{%
  }

\copyrightyear{2026}
\acmYear{2026}
\setcopyright{cc}
\setcctype{by}
\acmConference[SIGIR '26] {Proceedings of the 49th International ACM SIGIR Conference on Research and Development in Information Retrieval}{July 20--24, 2026}{Melbourne, VIC, Australia.}
\acmBooktitle{Proceedings of the 49th International ACM SIGIR Conference on Research and Development in Information Retrieval (SIGIR '26), July 20--24, 2026, Melbourne, VIC, Australia}
\acmISBN{979-8-4007-2599-9/2026/07}
\acmDOI{10.1145/3805712.3809733}

\newtcolorbox{promptbox}[1]{
  colback=white,
  colframe=gray,
  coltitle=white,
  fonttitle=\bfseries,
  title=#1,
  arc=4pt,
  boxrule=1.2pt,
  left=8pt,
  right=8pt,
  top=6pt,
  bottom=6pt
}

\begin{document}

\title{Bridging Behavior and Semantics for Time-aware Cross-Domain Sequential Recommendation}


\author{Zhida Qin}
\orcid{0000-0002-9270-1810}
\affiliation{%
  \institution{School of Computer Science, Beijing Institute of Technology}
  \city{Beijing}
  \country{China}}
\email{zanderqin@bit.edu.cn}

\author{Zemu Liu}
\orcid{0009-0000-6480-3118}
\affiliation{%
  \institution{School of Computer Science, Beijing Institute of Technology}
  \city{Beijing}
  \country{China}
}
\email{zemu\_liu@bit.edu.cn}

\author{Haoyan Fu}
\orcid{0009-0002-3263-2427}
\affiliation{%
  \institution{School of Computer Science, Beijing Institute of Technology}
  \city{Beijing}
  \country{China}
}
\email{haoyan-fu@bit.edu.cn}

\author{Chong Zhang}
\orcid{0000-0002-9763-3351}
\authornote{Corresponding author}
\affiliation{%
  \institution{School of Computer Science and Technology, Xi'an Jiaotong University}
  \city{Xi'an}
  \country{China}}
\email{zhangchong@xjtu.edu.cn}

\author{Tianyu Huang}
\orcid{0000-0003-2307-2261}
\affiliation{%
  \institution{School of Computer Science, Beijing Institute of Technology}
  \city{Beijing}
  \country{China}}
\email{huangtianyu@bit.edu.cn}

\author{Yidong Li}
\orcid{0000-0003-2965-6196}
\affiliation{%
  \institution{School of Computer Science and Technology, Beijing Jiaotong University}
  \city{Beijing}
  \country{China}}
\email{ydli@bjtu.edu.cn}

\author{Gangyi Ding}
\orcid{0009-0001-4185-4833}
\affiliation{%
  \institution{School of Computer Science, Beijing Institute of Technology}
  \city{Beijing}
  \country{China}}
\email{dgy@bit.edu.cn}

\renewcommand{\shortauthors}{Zhida Qin et al.}

\begin{abstract}
Cross-domain sequential recommendation (CDSR) alleviates interaction sparsity by jointly modeling user behaviors across multiple domains. While current studies have made some progresses, they still neglect two issues that severely impact recommendation performance: (i) \textit{ignoring domain-specific interaction frequencies and interest decay rates at identical time intervals;} (ii) \textit{treating semantic preferences as time-invariant during cross-domain transfer.} To address these, we propose a novel framework that bridges \textbf{B}ehavior and \textbf{S}emantics for \textbf{T}ime-aware \textbf{C}ross-\textbf{D}omain \textbf{S}equential \textbf{R}ecommendation (\textbf{BST-CDSR}). Specifically, we design a behavioral preference evolution module that decouples long-term interests and short-term intentions, and models continuous-time preference via a neural ordinary differential equation (ODE) with event-driven updates. Additionally, to capture time-aware semantic preferences, we introduce a temporal counterfactual-enhanced semantic generator that discretizes temporal interval tokens and leverages large language models (LLMs) to extract robust temporal semantics, where counterfactual perturbations enhance the time sensitivity of semantic preferences. Furthermore, we propose a time-preference guided domain transfer module to adaptively control transfer weights and mitigate negative transfer. Extensive experiments on real-world datasets demonstrate that BST-CDSR consistently outperforms baselines.
\end{abstract}




\maketitle

\section{INTRODUCTION}
With the rapid development of online platforms, users are exposed to various massive amounts of content \cite{sr1, sr2}. To improve the user's experience, recommendation systems have become an indispensable part \cite{qin1, qin2, qin3, qin4}. However, due to the exponential growth of internet content, user-item interactions are typically extremely sparse, which significantly decreases recommendation quality \cite{qin5, qin6, tgode}. In recent years, cross-domain sequential recommendation (CDSR) has gained increasing attention as a promising solution \cite{mifn, cmvcdr, iesrec}. By jointly modeling user sequences from multiple domains, CDSR provides richer information for capturing user preferences, thereby mitigating interaction sparsity and improving recommendation accuracy \cite{man, amid, psjnet, da-gcn}.

Typical CDSR studies primarily rely on ID information and focus on fully overlapping users, enabling models to explicitly capture behavioral correlations across domains \cite{cgrec,ca-cdsr,syncrec,hgncdsr}. The key challenge lies in integrating cross-domain knowledge from distinct distributions \cite{dmcdr, c2dsra2,cd-asr}. While effective, recommendations based only on user behaviors often yield unsatisfactory results. Recently, with the rapid advancement of large language models (LLMs), several studies have explored incorporating world knowledge from LLMs to inject semantic information into CDSR \cite{llm-edt, llmcdsr}. For instance, LLM4CDSR \cite{llm4cdsr} leverages LLMs to learn unified cross-domain semantic representations from item textual descriptions, while URLLM \cite{urllm} extracts item attribute semantics via LLMs to construct an item–attribute graph.


Despite their effectiveness, these approaches largely overlook the domain-dependent temporal characteristics of user interactions, thereby limiting their ability to accurately utilize cross-domain information. Specifically, they suffer from two major issues in modeling cross-domain interactions: (i) \textit{Ignoring domain-specific interaction frequencies and interest decay rates, where the same time interval may correspond to different preference states across different domains;} (ii) \textit{Treating semantic preferences as static features, failing to distinguish between temporary situational needs and stable user preferences.} Such omissions may prevent models from accurately capturing user behavior and semantic preferences, resulting in inefficient utilization of cross-domain information or even negative transfer. To address these issues, explicitly incorporating temporal factors into CDSR is essential. However, jointly modeling temporal characteristics and cross-domain interactions faces the following challenges:

\begin{itemize}[leftmargin=*, itemindent=0pt, nosep]
    \item \textbf{How to accurately model preference evolution under cross-domain temporal misalignment?} Real-world user behaviors exhibit high irregularity and domain-dependent temporal patterns, with interaction frequency and interval distributions varying significantly across domains. Taking the Movie-Book domain as an example, since watching a movie takes far less time than reading a book, over the same time scale (e.g., a week), users’ interaction intervals in the movie domain should be shorter than those in the book domain (as shown is Figure \ref{data-analysis}(a)), and users’ interests in the movie domain may have shifted several times, while their interests in the book domain may remain consistent. As a result, identical time intervals may induce different degrees of preference drift in different domains. This discrepancy becomes particularly critical in cross-domain scenarios, as mixed-domain sequences tend to obscure domain-specific temporal signals. Ignoring such domain-dependent temporal patterns may lead to inaccurate evolution of user preference. Therefore, the model should be able to capture domain-specific temporal preferences in continuous time, while remaining robust to non-uniform and heterogeneous time intervals across domains.
    \item \textbf{How to model time-dependent semantics to prevent negative transfer?} Semantic information plays a crucial role in bridging different domains; however, semantic preferences are inherently time-dependent rather than static. Some semantic interactions stem not from users' inherent preferences but from situational needs triggered by specific temporal contexts (e.g., holidays). For instance, users watching Christmas movies during Christmas may primarily be driven by the holiday atmosphere; once the holiday passes, the interest in such content typically vanishes. Treating semantic information as a static entity, while overlooking its temporal validity, risks misinterpreting such ‘situational semantics’ as ‘stable user preferences’. This misinterpretation is particularly detrimental in cross-domain scenarios, as context-specific semantics from the source domain may be erroneously transferred to the target domain, resulting in negative transfer. Consequently, models should capture the temporal dependencies of semantics to accurately model user preferences.
    
\end{itemize}

To address these challenges, we propose a novel framework that bridges \textbf{B}ehavior and \textbf{S}emantics for \textbf{T}ime-aware \textbf{C}ross-\textbf{D}omain \textbf{S}equential \textbf{R}ecommendation (\textbf{BST-CDSR}). First, we introduce a behavioral preference evolution module that decouples user behaviors into long-term interests and short-term intentions in different domains, modeling the preference evolution over continuous time via neural ODEs with event-driven updates. Then, we design a temporal counterfactual-enhanced semantic preference generator that converts time intervals into discrete tokens and leverages LLMs to extract time-aware semantic preferences. Additionally, we incorporate counterfactual perturbations of time gap tokens to enhance temporal of semantic preferences sensitivity to temporal shifts. Finally, we propose a time-preference guided domain transfer module that explicitly utilizes temporal patterns and preference relevance to adaptively control domain transfer strength, thereby mitigating negative transfer and enabling personalized cross-domain recommendations. Extensive experiments on real-world datasets demonstrate that BST-CDSR consistently outperforms strong baselines.

The key contributions of this paper are as follows:
\begin{itemize}[nosep]
\item We propose a novel CDSR framework that incorporates temporal information across three critical stages: behavioral preference modeling, semantic preference modeling, and cross-domain transfer.
\item We design a behavioral preference evolution module that decouples user preferences into long-term interests and short-term intentions, modeling the preference evolution process over continuous time.
\item We propose a temporal counterfactual-enhanced semantic preference generator that enhances the sensitivity of semantic preferences to time by discretizing time intervals and introducing counterfactual temporal perturbations.
\item We introduce a time-preference guided domain transfer mechanism that jointly models users' temporal patterns and preference relevance to adaptively assign domain transfer weights.
\end{itemize}

\section{DATA ANALYSIS}

In this section, we illustrate two critical issues prevalent in cross-domain sequential data, which are largely ignored by existing methods: (i) \textit{irregular and domain-dependent interaction intervals,} and (ii) \textit{the temporal validity of semantic preferences.} We utilize the Movie-Book dataset to visualize the statistical patterns.

\begin{figure}[t!]
    \centering
    \begin{minipage}{\columnwidth}
        \centering
        \begin{subfigure}[b]{0.49\columnwidth}
            \centering
            \includegraphics[width=\linewidth]{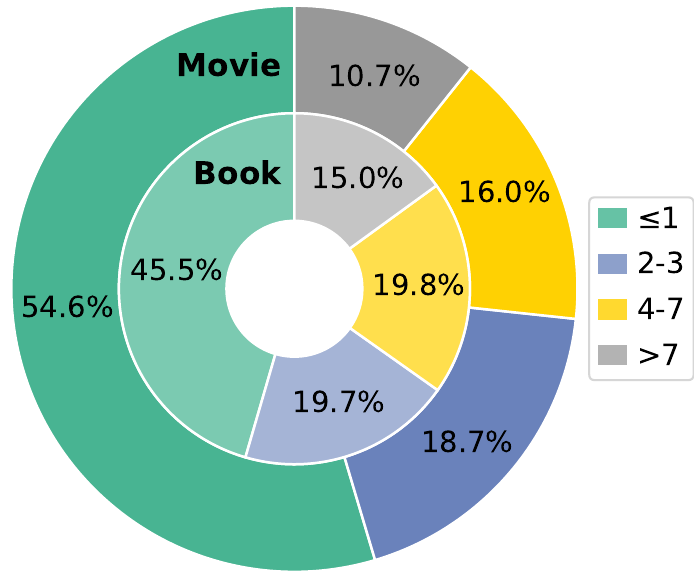}
            {\small (a) Time Interval Ratios}
        \end{subfigure}
        \hfill
        \begin{subfigure}[b]{0.49\columnwidth}
            \centering
            \includegraphics[width=\linewidth]{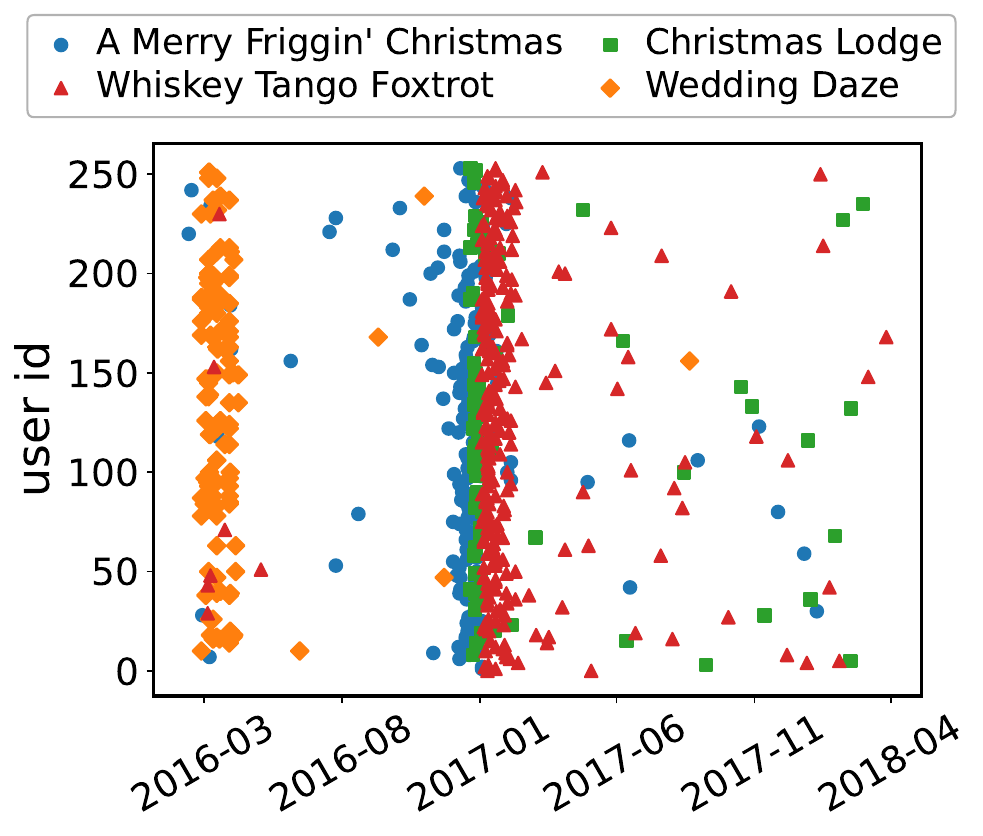}
            {\small (b) Semantic-Temporal Correlation}
        \end{subfigure}
        \vspace{-0.5em}
    \end{minipage}
    \vspace{-0.3em}
    \caption{Data Analysis on Behavior and Semantic. (a) Behavioral view: distribution of interaction time intervals across different domains; (b) Semantic view: interaction timelines of representative movie items with strong temporal correlations.}
    \label{data-analysis}
\end{figure}

We present the analysis results in Figure \ref{data-analysis} to demonstrate these issues clearly: Figure \ref{data-analysis}(a) displays the distribution of time intervals between neighboring interactions in different domains; Figure \ref{data-analysis}(b) visualizes the interaction timelines of representative movie items exhibiting strong semantic-temporal correlations. From these data analysis figures, we can observe the following two phenomena:

\textbf{Irregular interaction intervals and domain-specific interaction frequencies.} As shown in Figure \ref{data-analysis}(a), the time intervals between user interactions exhibit significant non-uniformity. Specifically, nearly half of the interactions are concentrated within a single day. While a substantial number of interactions occur at longer intervals, with 10\% spanning over a week, suggesting users may experience distinct preference shifts over time. Additionally, the movie domain exhibits approximately 10\% more interactions within a single day than the book domain, while displaying 5\% fewer interactions in long intervals. This indicates distinct interaction frequencies between domains, meaning the same time intervals may reflect different degrees of preference drift across domains. Therefore, ignoring domain-specific temporal patterns may fail to accurately model user preferences. This hinders accurate cross-domain transfer capabilities, ultimately resulting in suboptimal recommendation performance.

\begin{figure*}[t!]
    \centering
    \includegraphics[width=0.95\linewidth]{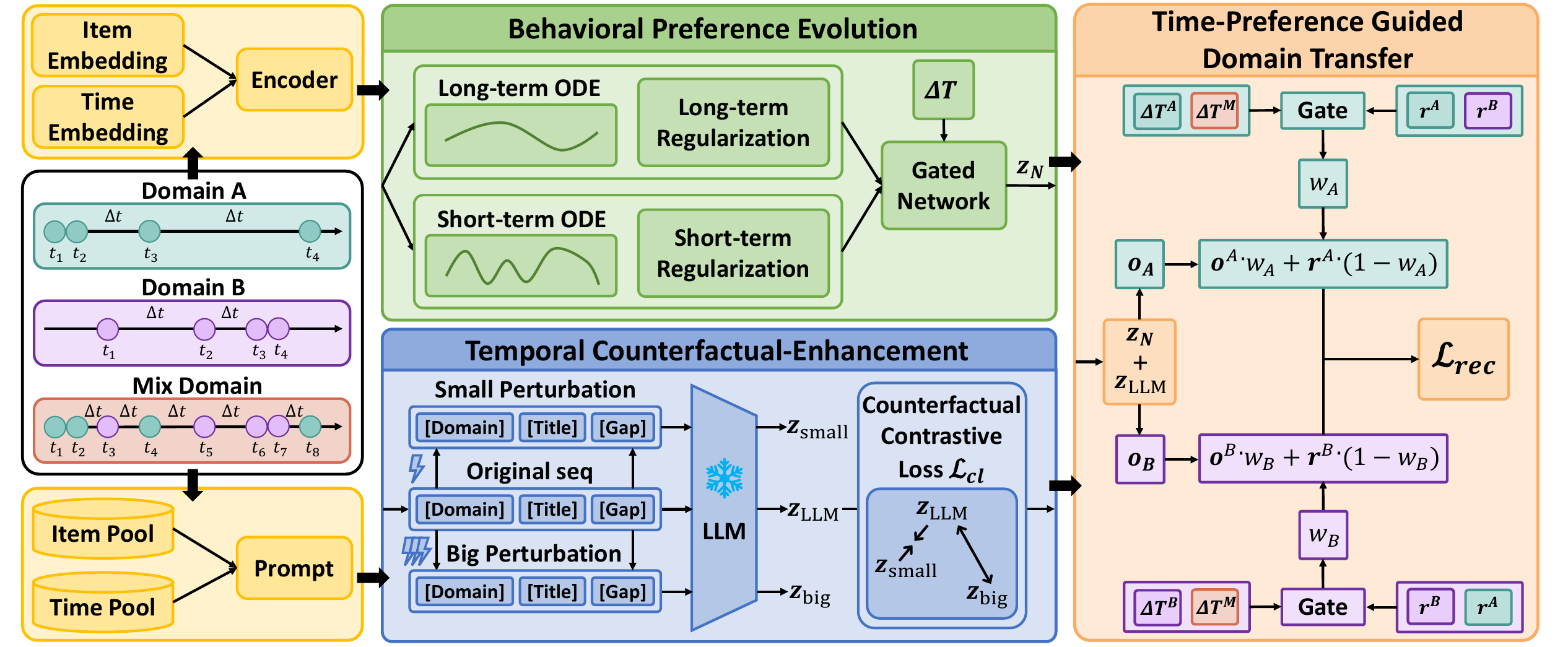}
    \caption{The overall architecture of the proposed method. The model consists of three key components: (i) a time-aware behavioral preference evolution module (green), (ii) a temporal counterfactual-enhanced semantic preference generator (blue), and (iii) a time-preference guided adaptive domain transfer (orange) for final recommendation.}
    \label{method}
\end{figure*}

\textbf{Strong correlation between item semantics and interaction time.} We observe that semantic preferences are strongly influenced by time. Figure \ref{data-analysis}(b) illustrates representative movie items whose interaction frequencies exhibit clear temporal clustering patterns consistent with their semantic. Such items can be broadly categorized into three representative types: those associated with specific holidays (e.g., Merry Christmas), seasonal themes (e.g., Wedding Fever), or major real-world events (e.g., Tango, War, Whiskey). These items typically exhibit peak interactions within specific time periods. These examples demonstrate that some semantic interactions are driven by situational contexts rather than user preferences. Consequently, modeling items as static semantic entities may preserve outdated situational semantics, resulting in noise rather than knowledge during cross-domain transfer and directly causing negative transfer.

The above analysis reveals that time carries rich information beyond simple ordering. Therefore, it is necessary to incorporate domain-dependent interaction intervals and semantic-temporal correlations into the model to better capture users' dynamic preferences and improve recommendation accuracy.

\section{PRELIMINARY}
We focus on a standard CDSR scenario involving two domains, denoted as domain $A$ and domain $B$. For a user $u$, each interaction is represented by an item–time pair $(i, t)$, where $i$ indicates the interacted item 
and $t$ records the occurrence time. Accordingly, the user's historical interactions in domain $A$ and domain $B$ can be described as two chronologically ordered sequences:
\begin{gather}
\mathcal{S}^{A} = \big\{(i_{k}^{A}, t_{k}^{A}) \mid i_{k}^{A} \in I^{A}, \, t_{k}^{A} \in T \big\}_{k=1}^{n}, \\
\mathcal{S}^{B} = \big\{(i_{k}^{B}, t_{k}^{B}) \mid i_{k}^{B} \in I^{B}, \, t_{k}^{B} \in T \big\}_{k=1}^{m},
\end{gather}
where $I^{A}$ and $I^{B}$ denote the item sets of the two domains, and $T$ represents the shared time space. To capture the temporal evolution of user preferences across domains and to facilitate the learning of domain-shared behavioral patterns, we further construct a unified 
cross-domain sequence $\mathcal{S}^{M}$ by merging $\mathcal{S}^{A}$ and $\mathcal{S}^{B}$ chronologically. Given the historical interactions, the task is to predict the user’s next item 
in each domain, i.e., $i_{n+1}^{A}$ and $i_{m+1}^{B}$.

\section{METHOD}
To address the two issues, we present the details of our model, as shown in Figure \ref{method}. First, we feed the interaction sequences from domain A, domain B, and the mixed domain into a \textbf{behavioral preference evolution} module to capture users’ time-dependent behavioral preferences under irregular intervals across domains. Next, we leverage a \textbf{temporal counterfactual-enhanced semantic preference generator} to extract time-aware semantic preferences. Finally, we fuse behavioral and semantic preferences for each domain, and pass the fused representations to a \textbf{time-preference guided domain transfer} module, which adaptively controls the transfer weights and produces the final recommendation.

\subsection{Embedding Layer}
In this section, we map raw item IDs and timestamps to a dense representation. Specifically, we separately model absolute time and relative time to capture more comprehensive temporal patterns.

\subsubsection{Item Encoding}
We maintain domain-specific item embedding tables $\boldsymbol{M}^{A}\in\mathbb{R}^{|I^{A}|\times d}$ and 
$\boldsymbol{M}^{B}\in\mathbb{R}^{|I^{B}|\times d}$, along with a domain-shared matrix 
$\boldsymbol{M}^{M}\in\mathbb{R}^{(|I^{A}|+|I^{B}|)\times d}$. Given an input item sequence, the corresponding item embeddings are retrieved via table lookup. Sequences shorter than length $N$ are padded with zero vectors.

\subsubsection{Temporal Encoding}
We model temporal information from two complementary perspectives: the absolute time of interactions, and the relative time between adjacent interactions.

For each timestamp $t_i$, we map it to a learnable embedding vector via a lookup table $\boldsymbol{M}^{at}\in\mathbb{R}^{|T|\times d}$, which is obtained by employing a mapping function $I: t_i\rightarrow\{1,2,\ldots,|T|\}$.

In order to obtain the relative time embedding, we first calculate the time gaps $\Delta_i = t_i - t_{i-1}$, with $\Delta_1$ assigned -1 to mark sequence initiation. Each gap is discretized by using a logarithmic transformation: $\text{pos}_i=\lfloor a\log(\Delta_i+2)\rfloor$, and embedded via a lookup table 
$\boldsymbol{M}^{rt}\in\mathbb{R}^{|\mathcal{P}|\times d}$, where $|\mathcal{P}|$ is the number of discretized intervals.

\subsection{Behavioral Preference Evolution}
Existing methods often assume interactions are uniformly distributed, neglecting \textbf{temporal irregularity} and \textbf{domain-dependent interaction intervals}. As a result, the same time intervals may correspond to different degrees of cross-domain preference drift. More critically, user preferences \textbf{continuously evolve between interactions}, while discrete sequence encoders update only at interaction points, making it difficult to capture this gradual evolution and the cumulative impact of long intervals. To address this, we decouple preferences into long-term and short-term states, modeling their evolution over continuous time to simultaneously capture gradual interest drift and sharp behavioral shifts. This process is independently applied across three domains, with domain subscripts omitted for brevity.

\subsubsection{ODE-based Preference Modeling.}
Given a user interaction sequence $\mathcal{S} = \left((i_{1},t_1), (i_{2}, t_2) \ldots, (i_{N}, t_N)\right)$, we first employ a Transformer layer to capture item-level preference representations $\boldsymbol{E} = \text{Att}\left(\text{Emb}(\mathcal{S})\right)$. Each $\boldsymbol{e}_t \in \boldsymbol{E}$ encodes the user’s discrete instantaneous preference at the $t$-th interaction event.

\textbf{Long-term Preference Modeling.} We formulate user preference evolution as a continuous-time dynamical process augmented by event-driven updates: between interactions, user preferences evolve smoothly over continuous time; when an interaction occurs, preferences exhibit discrete shifts driven by the current behavior.

Specifically, we maintain the long-term state $\boldsymbol{h}^L_t \in \mathbb{R}^d$ at each time step $t$. Between consecutive interactions, we evolve long-term state smoothly under a neural ODE:
\begin{equation}
    \frac{d\boldsymbol{h}^L_t}{dt} = f_L(\boldsymbol{h}^L_t, \tilde{\Delta}_t), \quad
\end{equation}
where $\tilde{\Delta}_t = \log_2(\Delta_t + 2) / \log_2(\Delta_{\text{max}} + 2)$ denotes the normalized interval, $f_L$ is parameterized by a neural network, such as MLPs. To obtain the pre-interaction state ${\boldsymbol{h}_t^{L}}^\prime$ before the $(t+1)$-th interaction, we solve the above ODE over the interval $\tilde{\Delta}_t$. In practice, we employ a first-order Euler approximation for numerical integration:
\begin{equation}
    {\boldsymbol{h}_t^{L}}^\prime = \boldsymbol{h}^L_t + \tilde{\Delta}_t \cdot f_L(\boldsymbol{h}^L_t, \tilde{\Delta}_t).
\end{equation}

However, the continuous ODE evolution only models between-interaction preference drift and cannot explicitly incorporate the event-driven changes triggered by user interactions. Therefore, upon observing interaction $i_{t+1}$, we update the long-term state using the transient interaction $\boldsymbol{e}_{t+1}$:
\begin{equation}
    \boldsymbol{h}^L_{t+1} = \text{GRU}(\boldsymbol{e}_{t+1}, {\boldsymbol{h}_t^{L}}^\prime).
\end{equation}

\textbf{Short-term Preference Modeling.}
We model short-term preferences $\boldsymbol{h}^S_t \in \mathbb{R}^d$ using the same continuous-time evolution and event-driven update as the long-term state, but with independent parameters to capture rapidly changing, transient user intentions.

\textbf{Preferences Fusion.}
Finally, we design a time-aware gated network that fuses long-term and short-term states to obtain the user's behavioral preferences at each time step. This adaptive fusion enables the model to dynamically adjust long- and short-term dependencies based on the duration of time intervals, thereby achieving an accurate, time-aware representation of user behavioral preferences. The formulas are expressed as:
\begin{gather}
    \boldsymbol{g}_{t+1} = \sigma \,(\boldsymbol{W}[\boldsymbol{h}^L_{t+1}, \boldsymbol{h}^S_{t+1}, \tilde{\Delta} _t]), \\
    \boldsymbol{z}_{t+1} = \boldsymbol{g}_{t+1} \odot \boldsymbol{h}^S_{t+1} + (1 - \boldsymbol{g}_{t+1}) \odot \boldsymbol{h}^L_{t+1},
\end{gather}
where $\sigma(\cdot)$ denote the sigmoid function.

We will perform such process for each time step in chronological order to obtain the sequence preference $\boldsymbol{Z} \in \mathbb{R}^{N \times d}$. For domain $A$ and domain $B$, we take the last step representation $\boldsymbol{z}_{N}$ as the user's behavioral preference. For the first time step, we set $\boldsymbol{h}^L_1 = \boldsymbol{h}^S_1 = \boldsymbol{e}_1$.

\subsubsection{Temporal Regularization.}
Although the two-behavior neural ODE encoder can comprehensively model the evolution of user behavioral preferences, unconstrained neural networks may produce the opposite effect, blurring the distinction between long-term and short-term states. To further enhance their expected role, we introduce two temporal regularization objectives to constrain the evolution of $\boldsymbol{h}^L_t$ and $\boldsymbol{h}^S_t$.

\textbf{Long-term Regularization.} Long-term preferences are expected to evolve smoothly over time, reflecting persistent user interests rather than abrupt fluctuations. Therefore, we apply temporal smoothing regularization to the long-term states $\boldsymbol{h}^L_t$:
\begin{equation}
    \mathcal{L}_L=\sum_{t=2}^N (1-\tilde{\Delta}_{t})\left\|\boldsymbol{h}^L_t-\boldsymbol{h}^L_{t-1}\right\|_2^2,
\end{equation}
where $1-\tilde{\Delta}_{t}$ is a time-dependent weight that adjusts the constraint strength based on the time interval. This design applies stronger constraints to short intervals and weaker constraints to long intervals, enabling the long-term state to accurately capture the user's interest drift.

\textbf{Short-term Regularization.} Short-term states are designed to closely track the user’s immediate behavioral preference at each interaction. Hence, we introduce a contrastive alignment objective between $\boldsymbol{h}^S_t$ and $\boldsymbol{e}_t$. Specifically, we encourage $\boldsymbol{h}^S_t$ to be maximally similar to its corresponding $\boldsymbol{e}_t$ while maintaining distinction from other interactions in the same batch:
\begin{equation}
    \mathcal{L}_S=-\sum_t\log\frac{\exp(\text{sim}(\boldsymbol{h}^S_t,\boldsymbol{e}_t)/\tau)}{\sum_j\exp(\text{sim}(\boldsymbol{h}^S_t,\boldsymbol{e}_j)/\tau)}.
\end{equation}

These two temporal regularization jointly decouple long-term and short-term states, enabling neural ODEs to model accurate user behavioral preferences under irregular temporal interactions.

\subsection{Temporal Counterfactual-Enhanced Semantic Preference Generator}
Recent studies increasingly incorporate semantics as complementary information to behavioral signals in CDSR, but most methods model semantics as static entities, neglecting that semantic preferences are influenced by both \textbf{situational contexts} and \textbf{user preferences}. Furthermore, prior studies \cite{llmweak1, llmweak2} indicate that LLMs exhibit \textbf{limited ability} to reason over absolute numerical values, making it difficult to directly model exact interaction timestamps. Therefore, we divide exact time intervals into discrete tokens and introduce counterfactual enhancement, enabling the model to capture accurate time-aware semantic preference. This process is independently applied across three domains, with domain subscripts omitted for brevity.

\subsubsection{Time-Aware Semantic Preference Extraction.}
To model users' semantic preferences, we employ LLMs to encode interactions augmented with explicit temporal tokens. We feed entire interaction sequences as structured textual input, enabling LLMs to reason jointly about both semantic content and temporal patterns. 

\textbf{Discrete-Time-Based Prompt Construction.}
Given a user interaction sequence $\mathcal{S}$ and timestamp interval sequence $\Delta T = (\Delta_1,\Delta_2,\ldots,\Delta_N)$, we can first obtain the textual representation $\text{[Title]}_i$ of the items. Since LLMs exhibit limited reasoning of continuous numerical time, we map $\Delta_i$ to discrete time interval token:
\begin{equation}
    \text{[Gap]}_j = \phi(\Delta_i), \quad \text{[Gap]}_j \in \mathcal{V}_{\text{gap}},
\end{equation}
where $\phi(\cdot)$ denotes the discretization function, and $\mathcal{V}_{\text{gap}}$ represents a predefined set of time-gap tokens. Additionally, we prepend a domain token (e.g., $\text{[A\_Domain]}$ or $\text{[B\_Domain]}$) ahead of each interaction text, enabling the LLM to recognize domain transfers during interactions. Finally, by including the base prompt, we construct the following sequence of input tokens $\textbf{x}$:

\begin{promptbox}{Prompt Template}
You will act as a time-aware preference interpreter. Please extract the time-aware semantic preferences from the following interaction sequence: \\
$\text{[Domain]}\:\text{[Title]}\:\text{[Gap]}\:\text{[Domain]}\:\text{[Title]}\:\ldots$
\end{promptbox}

\textbf{Semantic Preference Extraction.}
Input the tokenized sequence $\textbf{x}$ into the LLM:
\begin{equation}
    \boldsymbol{Z}_{\text{LLM}} = \text{LLM}(\textbf{x}), \quad \boldsymbol{Z}_{\text{LLM}} \in \mathbb{R}^{L \times D_{\text{LLM}}},
\end{equation}
where $L$ denotes the token length and $D_{\text{LLM}}$ denotes the LLM embedding dimension. We take the hidden state corresponding to the last valid token as the user's time-aware semantic representation. As the semantic space of the LLM differs from the latent space of the recommendation model, we employ the LLMEmb \cite{llmemb} method to reduce its dimension to the model dimension $d$:
\begin{gather}
    \boldsymbol{z}_{\text{LLM}} = f_{\text{Adapt}}(f_{\text{PCA}}( \boldsymbol{Z}_{\text{LLM}}[L])), \\
    f_{\text{PCA}}(\cdot):\mathbb{R}^{D_{\text{LLM}}}\to\mathbb{R}^{D_{\text{Mid}}}, \quad
    f_{\text{Adapt}}(\cdot):\mathbb{R}^{D_{\text{Mid}}}\to\mathbb{R}^d,
\end{gather}
where $\boldsymbol{z}_{\text{LLM}}$ denotes the obtained time-aware semantic preference.

\subsubsection{Counterfactual-based Time-Sensitivity Enhancement.} 
Though the above procedure enables LLMs to model time-aware semantic preferences, it suffer from insufficient sensitivity to preference shifts caused by different temporal patterns. To address this issue, we design a counterfactual enhancement strategy that disrupts temporal patterns while preserving the interaction semantics, thereby enabling the model to be more sensitive to temporal information.

\textbf{Construction of Counterfactual Sequences.}
We first partition the time-gap token set $\mathcal{V}_{\text{gap}}$ into three duration-based groups, where tokens within the same group represent similar intervals. Based on the original sequence, we then construct counterfactual sequences by perturbing gap tokens in two ways: \textit{small-range} and \textit{large-range}. For small-range perturbation, each gap token is replaced with another token sampled from the same group with probability $\alpha_1$, modeling minor timing fluctuations. For large-range perturbation, each token is replaced with a token sampled from a different group with probability $\alpha_2$, yielding substantially different temporal patterns that mimic distinct interaction styles. The perturbation operators are defined as:

\begin{equation}
    \mathbf{x}_{\text{small}} = \mathcal{T}_{\mathrm{small}}\left(\mathbf{x}\right), \quad 
    \mathbf{x}_{\text{big}} = \mathcal{T}_{\mathrm{big}}\left(\mathbf{x}\right),
\end{equation}
where $\mathcal{T}_{\mathrm{small}}$ and $\mathcal{T}_{\mathrm{big}}$ denote the small and large perturbation operators respectively. We apply the same procedure in Section 4.3.1 on the two counterfactual sequences to obtain the corresponding semantic preferences $\boldsymbol{z}_{\text{small}}$ and $\boldsymbol{z}_{\text{big}}$.

\textbf{Counterfactual Optimization Objective.}
We propose a ranking-based counterfactual objective to achieve temporal semantic consistency: compared to $\boldsymbol{z}_{\text{big}}$, $\boldsymbol{z}_{\text{LLM}}$ should be closer to $\boldsymbol{z}_{\text{small}}$; while $\boldsymbol{z}_{\text{big}}$ should be more similar to $\boldsymbol{z}_{\text{LLM}}$ than other users' representations within the same batch. The formulation can be expressed as:
\begin{gather}
    \mathcal{L}_{cl} = \mathcal{L}_1 + \mathcal{L}_2, \\
    \mathcal{L}_1=-\log \, \sigma \left(\left(\text{sim}(\boldsymbol{z}_{\text{LLM}},\boldsymbol{z}_{\text{small}})-\text{sim}(\boldsymbol{z}_{\text{LLM}},\boldsymbol{z}_{\text{big}})\right)/\tau \right), \\
    \mathcal{L}_2=-\log \, \sigma \left(\left(\text{sim}(\boldsymbol{z}_{\text{LLM}},\boldsymbol{z}_{\text{big}})-\text{sim}(\boldsymbol{z}_{\text{LLM}},\boldsymbol{z}_{\text{neg}})\right)/\tau \right), \\
    \text{sim}(\boldsymbol{z}_\text{LLM},\boldsymbol{z}_\text{neg})=\frac{1}{K}\sum_{j\in\mathcal{N}_K(u)}\text{sim}(\boldsymbol{z}^u_\text{LLM},\boldsymbol{z}^j_\text{LLM}),
\end{gather}
where $\mathcal{N}_K(u)$ denotes the top $K$ most similar negatives within the batch, and $\tau$ is the temperature coefficient. This objective encourages the model to be robust to minor temporal perturbations while remaining capable of distinguishing significant changes in temporal patterns.

\subsection{Time-Preference Guided Domain Transfer}
Most existing CDSR methods assume users exhibit identical dependency patterns across domains. However, user behaviors exhibit \textbf{high heterogeneity} in both temporal structure and cross-domain interactions. For instance, some users frequently switch domains within short intervals, while others focus on a single domain for extended periods. Ignoring such user heterogeneity will result in suboptimal transfer strategies and may even introduce negative transfer.

To address this issue, we propose a time-preference guided domain transfer module that explicitly models user heterogeneity. By leveraging temporal patterns and preference relevance, the model determines when to perform cross-domain transfer and how strongly to transfer. Therefore, it dynamically adjusts transfer weights at the user level, enabling personalized and adaptive cross-domain recommendation.

\textbf{Time-based Transfer Factor.}
Given a user's mixed domain sequence $\mathcal{S}^M$, each interaction is assigned a domain token $\textit{dom}$ based on its origin, which is then combined with the corresponding time interval $\Delta_t^M$ to construct the domain-aware temporal pattern sequence:
\begin{equation}
    {\Delta T}^M = \big\{(\text{dom}_i, \Delta_{t_i}^M) \mid \text{dom}_i \in \{A,B\} \big\}_{i=1}^{N}.
\end{equation}

Similarly, we can obtain the user's sequences ${\Delta T}^A$, ${\Delta T}^B$. These sequences explicitly encode how users switch between domains over time. We then encode them with an attention:
\begin{equation}
    \boldsymbol{u}^A_t = \text{Att}({\Delta T}^A), \quad
    \boldsymbol{u}^B_t = \text{Att}({\Delta T}^B), \quad
    \boldsymbol{u}^M_t = \text{Att}({\Delta T}^M).
\end{equation}

\textbf{Preference-based Transfer Factor.}
Simply relying on temporal patterns is insufficient to fully capture cross-domain dependencies, thus we utilize preference representations as a complement. Since the mixed domain captures domain-shared preferences, we utilize the mixed domain representation to reflect the preference correlation:
\begin{gather}
    \boldsymbol{r}^A = \text{AvgPool}\{ \boldsymbol{z}^M_k \mid \boldsymbol{z}_k^M \in \boldsymbol{Z}^M, \, i_k \in I^A\} + \boldsymbol{z}_{\text{LLM}}^M, \\
    \boldsymbol{r}^B = \text{AvgPool}\{ \boldsymbol{z}^M_k \mid \boldsymbol{z}_k^M \in \boldsymbol{Z}^M, \, i_k \in I^B\} + \boldsymbol{z}_{\text{LLM}}^M.
\end{gather}

\textbf{Time-Preference Hybrid Weights.}
Finally, we aggregate temporal patterns and preference correlations, and utilize a gated network to calculate personalized transfer weights:
\begin{gather}
    w_A = \sigma\left(f_{\text{gate}}\left(\left[\boldsymbol{W}_t^A[\boldsymbol{u}^A_t, \boldsymbol{u}^M_t], \boldsymbol{W}_r^A[\boldsymbol{r}^A, \boldsymbol{r}^B]\right]\right) \right), \\
    w_B = \sigma\left(f_{\text{gate}}\left(\left[\boldsymbol{W}_t^B[\boldsymbol{u}^B_t, \boldsymbol{u}^M_t], \boldsymbol{W}_r^B[\boldsymbol{r}^B, \boldsymbol{r}^A]\right]\right) \right), 
\end{gather}
where $\sigma(\cdot)$ denotes the sigmoid function. This design effectively reduces negative transfer, making the model adaptable to various cross-domain interaction temporal patterns and leading to more accurate and robust cross-domain transfer.

\subsection{Prediction and Optimization}
For a user $u$, we fuse behavioral preferences $\boldsymbol{z}^A_{N}$, $\boldsymbol{z}^B_N$ and semantic preferences $\boldsymbol{z}^A_{\text{LLM}}$, $\boldsymbol{z}^B_{\text{LLM}}$ by summation to obtain behavioral–semantic representations $\boldsymbol{o}^A$, $\boldsymbol{o}^B$. $\boldsymbol{r}^A$, $\boldsymbol{r}^B$ denote the cross-domain preference representations. By multiplying them with the adaptive weights $w_A$, $w_B$, we can predict the target items:
\begin{gather}
    \hat{\boldsymbol{y}}^{A} = \text{softmax}\left( ( w^{A} \boldsymbol{o}^{A} + (1 - w^{A}) \boldsymbol{r}^{A} ) \boldsymbol{W}^{A} \right), \\
    \hat{\boldsymbol{y}}^{B} = \text{softmax}\left( ( w^{B} \boldsymbol{o}^{B} + (1 - w^{B}) \boldsymbol{r}^{B} ) \boldsymbol{W}^{B} \right),
\end{gather}
where $\boldsymbol{W}^{A} \in \mathbb{R}^{d \times |I^A|}$, $\boldsymbol{W}^{B} \in \mathbb{R}^{d \times |I^B|}$ are the projection matrices and $\hat{\boldsymbol{y}}^{A}$, $\hat{\boldsymbol{y}}^{B}$ denote the probability of interacting with each item. For domains A and B, we employ the cross-entropy loss for optimization: 
\begin{gather}
    \mathcal{L}_A=-\frac{1}{|\mathcal{B}|}\sum_{u\in\mathcal{B}}\log\hat{\boldsymbol{y}}_{u,i_u}^A,\quad
    \mathcal{L}_B=-\frac{1}{|\mathcal{B}|}\sum_{u\in\mathcal{B}}\log\hat{\boldsymbol{y}}_{u,i_u}^B, \\
    \mathcal{L}_{\text{main}} = \mathcal{L}_A + \mathcal{L}_B,
\end{gather}
where $i_u$ is the ground-truth next item. Additionally, we introduce two types of constraints for each domain: temporal regularization for long- and short-term preferences, and counterfactual-enhanced temporal contrastive loss, formulated as:
\begin{gather}
    \mathcal{L}_{\text{ODE}} = \mathcal{L}_{\text{ODE}}^A + \mathcal{L}_{\text{ODE}}^B + \mathcal{L}_{\text{ODE}}^M, \quad \mathcal{L}_{\text{ODE}}^{\mathcal{D}} = \mathcal{L}_{L}^{\mathcal{D}} + \mathcal{L}_{S}^{\mathcal{D}}, \\
    \mathcal{L}_{\text{sem}} = \mathcal{L}_{cl}^A + \mathcal{L}_{cl}^B + \mathcal{L}_{cl}^M,
\end{gather}
where ${\mathcal{D}} \in \{ A, B,M\}$ denotes the domain placeholder. Finally, the complete optimization objective is:
\begin{equation}
    \mathcal{L} = \mathcal{L}_{\text{main}} + \lambda_1 \cdot \mathcal{L}_{\text{ODE}} + \lambda_2 \cdot \mathcal{L}_{\text{sem}},
\end{equation}
where $\lambda_1$, $\lambda_2$ denote the hyper-parameters of weights.


\section{EXPERIMENTS}

In this section, we perform comprehensive experiments on three real-world datasets to investigate the following research questions:
\begin{itemize}
\item \textbf{RQ1:} Is BST-CDSR able to maintain superior performance across all domains?
\item \textbf{RQ2:} What is the contribution of each component in BST-CDSR to the overall performance?
\item \textbf{RQ3:} Does the ODE module accurately model users' time-dependent preferences?
\item \textbf{RQ4:} Does the counterfactual module enhance temporal sensitivity in semantic preferences?
\item \textbf{RQ5:} Does it accurately decouple long- and short-term behavioral preferences?
\item \textbf{RQ6:} Does BST-CDSR alleviate the issue of negative transfer?
\end{itemize}

\subsection{Experimental Settings}
\subsubsection{Datasets.}
Extensive experimental evaluations are conducted on three benchmark datasets for CDSR tasks: (i) Food-Kitchen, (ii) Movie-Book, and (iii) Art-Office. The dataset classification follows the protocol introduced in $\text{C}^2\text{DSR}$ \cite{c2dsr}. The detailed dataset statistics are reported in Table \ref{dataset}.

\begin{table}[!t]
\setlength{\tabcolsep}{3pt}  
\small  
\renewcommand{\arraystretch}{1.0}  
\centering
\caption{Data statistics for three datasets.}
\vspace{-4mm}
\begin{tabular}{c|cccccc}
\toprule
\textbf{Datasets} & \textbf{\#Items} & \textbf{\#Train} & \textbf{\#Valid} & \textbf{\#Test} & \textbf{Avg.length} & \textbf{Sparsity} \\
\midrule
Food & 29,207 & \multirow{2}{*}{34,117} & \multirow{2}{*}{8,173} & \multirow{2}{*}{8,406} & 6.2 & 99.94\% \\
Kitchen & 34,886 & & & & 6.0 & 99.95\% \\
\midrule
Movie & 36,845 & \multirow{2}{*}{58,515} & \multirow{2}{*}{7,644} & \multirow{2}{*}{7,708} & 6.4 & 99.92\% \\
Book & 63,937 & & & & 5.6 & 99.96\% \\
\midrule
Art & 47,365 & \multirow{2}{*}{37,654} & \multirow{2}{*}{6,251} & \multirow{2}{*}{6,199} & 5.6 & 99.97\% \\
Office & 47,868 & & & & 5.8 & 99.96\% \\
\bottomrule
\end{tabular}
\label{dataset}
\end{table}

\input{baseline}

\subsubsection{Baselines.}
To evaluate the performance of BST-CDSR, we compare it with the following three types of baseline models: 

\textbf{i) the single-domain methods}:
\begin{itemize}[nosep]
\item \textbf{SASRec} \cite{sasrec} first introduces self-attention into sequential recommendation, effectively modeling user preferences.
\item \textbf{TiSASRec} \cite{tisasrec} extends SASRec by incorporating temporal information into the self-attention.
\item \textbf{CL4SRec} \cite{cl4srec} conducts three types of sequence-level contrastive learning to enhance recommendation performance.
\item \textbf{IOCRec} \cite{iocrec} designs a contrastive learning framework based on modeling users' intentions.
\end{itemize}

\textbf{ii) the CDSR methods}:
\begin{itemize}[nosep]
\item \textbf{$\text{C}^2\text{DSR}$} \cite{c2dsr} proposes a contrastive learning framework based on maximizing mutual information.
\item \textbf{EA-GCL} \cite{ea-gcl} designs an external attention encoder to address distribution bias in cross-domain interactions.
\item \textbf{Tri-CDR} \cite{tri-cdr} utilizes contrastive learning to model users' multi-granularity interests.
\item \textbf{ABXI} \cite{abxi} uses LoRa to model user preferences, addressing alignment issues across different domains.
\end{itemize}

\textbf{iii) the LLM-based CDSR methods}:
\begin{itemize}[nosep]
\item \textbf{LLMCDSR} \cite{llmcdsr} generates cross-domain pseudo-interactions via LLM and uses contrastive learning to address the issue of non-overlapping users.
\item \textbf{LLM4CDSR} \cite{llm4cdsr} leverages LLMs to obtain unified embedding representations and uses LLMs to generalize user preferences.
\item \textbf{LLM-EDT} \cite{llm-edt} introduces a two-stage training strategy to address the issue of domain imbalance.
\end{itemize}

\subsubsection{Evaluation Protocol and Implementation Details.}
We follow the common experimental protocol by partitioning the data into training, validation, and test sets with a ratio of 8:1:1. For evaluation, each test instance is associated with one positive sample and 999 randomly selected negative samples.

For a fair comparison, all methods share the same hyperparameter settings: the embedding dimension and batch size are both set to 256, the maximum number of training epochs is 100. In addition, we employ Llama2-7B for the LLM-based baselines. For BST-CDSR, we set the learning rate to 0.0005 with the Adam optimizer. We set $\lambda_1$ to 0.01 and $\lambda_2$ to 0.001. All experiments are conducted on a single NVIDIA GeForce RTX 3090 GPU (24GB). 

\begin{table}[!t]
\centering
\caption{Ablation Study of the proposed components on two datasets.}
\vspace{-4mm}
\small
\setlength{\tabcolsep}{3pt} 
\renewcommand{\arraystretch}{1.0}

\begin{tabular}{c|ccc|ccc}
\toprule
\toprule
\multirow{2}{*}{\textbf{Variants}}&\multicolumn{3}{c}{\textbf{Food}}&\multicolumn{3}{c}{\textbf{Kitchen}} \\ 
&{MRR}&{N@10}&{H@10}&{MRR}&{N@10}&{H@10}\\
\midrule
Single-ODE &0.1687&0.1731&0.2066&0.1106&0.1120&0.1388 \\
Dual LS-ODE &0.1712&0.1753&0.2088&0.1120&0.1139&0.1406 \\
+Semantic &0.1887&0.1959&0.2457&0.1186&0.1222&0.1566 \\
+CF-Enhance &0.1890&0.1965&0.2466&0.1196&0.1227&0.1571 \\
Exact-time &0.1935&0.2015&0.2516&0.1204&0.1233&0.1570 \\
BST-CDSR &\textbf{0.1949}&\textbf{0.2028}&\textbf{0.2547}&\textbf{0.1213}&\textbf{0.1246}&\textbf{0.1585} \\
\toprule
\multirow{2}{*}{\textbf{Variants}}&\multicolumn{3}{c}{\textbf{Movie}}&\multicolumn{3}{c}{\textbf{Book}} \\ 
&{MRR}&{N@10}&{H@10}&{MRR}&{N@10}&{H@10}\\
\midrule
Single-ODE &0.1236&0.1254&0.1532&0.0950&0.0943&0.1069 \\
Dual LS-ODE &0.1271&0.1285&0.1562&0.0964&0.0964&0.1114 \\
+Semantic &0.1329&0.1368&0.1814&0.1091&0.1123&0.1436 \\
+CF-Enhance &0.1374&0.1418&0.1876&0.1100&0.1126&0.1447 \\
Exact-time &0.1368&0.1423&0.1832&0.1095&0.1120&0.1439 \\
BST-CDSR &\textbf{0.1383}&\textbf{0.1436}&\textbf{0.1928}&\textbf{0.1123}&\textbf{0.1147}&\textbf{0.1457} \\
\bottomrule
\bottomrule
\end{tabular}
\label{ablation}
\end{table}

\subsection{Performance Comparison (RQ1)}
In this section, we conduct an comprehensive comparison between our method and several representative baselines. The experimental results on the three datasets are summarized in Table \ref{baseline}, from which the following conclusions can be drawn:

\textbf{BST-CDSR consistently outperforms other methods across all domains.} BST-CDSR achieves superior performance in all six domains spanning the three datasets, with noticeable improvements over existing approaches. This performance gain can be largely attributed to the effective collaboration of its three temporal components. Specifically, the behavioral preference evolution and counterfactual enhancement module accurately captures users' time-dependent behavioral and semantic preferences, while the domain transfer module adaptively merges cross-domain information. These results indicate that introducing temporal information is crucial for improving the performance of cross-domain sequential recommendation.

\textbf{Cross-domain information is effective in improving recommendation performance.} Single-domain SR methods generally perform worse than cross-domain methods because they fail to ease interaction sparsity. In contrast, cross-domain methods leverage information from source domains to effectively mitigate data sparsity and enhance recommendation accuracy. Notably, since IOCRec models users' multi-intentions, it performs close to some CDSR models.

\textbf{Semantic information can significantly increase recommendation performance.} Compared to models lacking semantic information, LLM-based models demonstrate a remarkable improvement in performance, as they are able to capture rich item semantics and high-level contextual relationships beyond interaction signals. This suggests that incorporating semantic representations can effectively complement behavioral information, leading to more accurate recommendations.

\subsection{Ablation Study (RQ2)}

In this section, in order to evaluate the impact of each module in BST-CDSR on performance, we design the following variants:

\begin{itemize}
\item \textbf{V1 Single-ODE:} Based on a layer of transformer, add a neural ODE which is not decoupled to  long- and short-term behavior.
\item \textbf{V2 Dual LS-ODE:} Replace the single ODE in V1 with a dual-behavior neural ODE based on short- and long-term.
\item \textbf{V3 +Semantic:} Adding semantic preference modeling on top of V2.
\item \textbf{V4 +CF-Enhance:} Introduce the temporal counterfactual-enhanced module based on V3.
\item \textbf{V5 Exact-time:} Replace the time-gap tokens in the semantic sequence with exact time, based on the complete model.
\end{itemize}

The comparative results of BST-CDSR and all its variants are shown in Table \ref{ablation}, from which we can observe:

First, the performance consistently improves as we progressively add modules, indicating that each component makes a positive contribution to BST-CDSR. Replacing \textbf{Single-ODE} with \textbf{Dual LS-ODE} yields steady gains, demonstrating the necessity of explicitly decoupling user behaviors into long- and short-term interests. This validates our behavioral preference evolution module can better captures preference evolution under irregular time intervals.

Besides, comparing \textbf{+CF-Enhance} with \textbf{+Semantic} further verifies the effectiveness of the temporal counterfactual-enhanced module. By perturbing time-gap tokens with both small- and large-range changes and employing temporal ranking contrastive objectives, the model can be more sensitive to temporal patterns while remaining robust to minor temporal fluctuations.

Moreover, BST-CDSR surpasses \textbf{+CF-Enhance}, highlighting the importance of the time-preference guided domain transfer module, which further improves performance by adaptively controlling cross-domain transfer weights according to both temporal patterns and preference correlations.

BST-CDSR outperforms \textbf{Exact-time} on all datasets, which supports our design of using discretized time-gap tokens rather than injecting exact numerical timestamps into the LLM. This validates that LLMs are less effective at reasoning over absolute numerical values, and that discrete time tokens provide a more LLM-friendly way to model time-aware semantic preferences.

\begin{figure}[t!]
    \centering
    \includegraphics[width=\columnwidth]{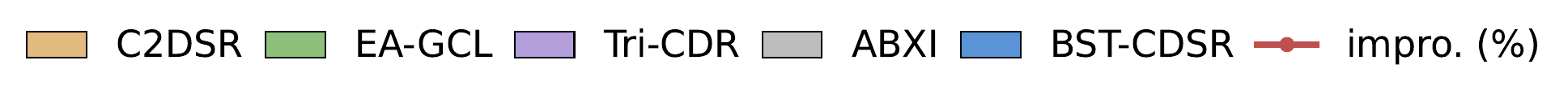}
    \vspace{0.3em}
    \begin{minipage}{\columnwidth}
        \centering
        \begin{subfigure}[b]{0.49\columnwidth}
            \centering
            \includegraphics[width=\linewidth]{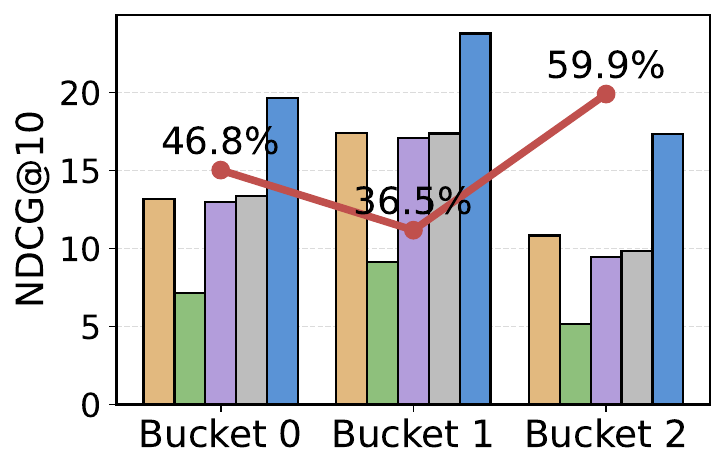}
        \end{subfigure}
        \hfill
        \begin{subfigure}[b]{0.49\columnwidth}
            \centering
            \includegraphics[width=\linewidth]{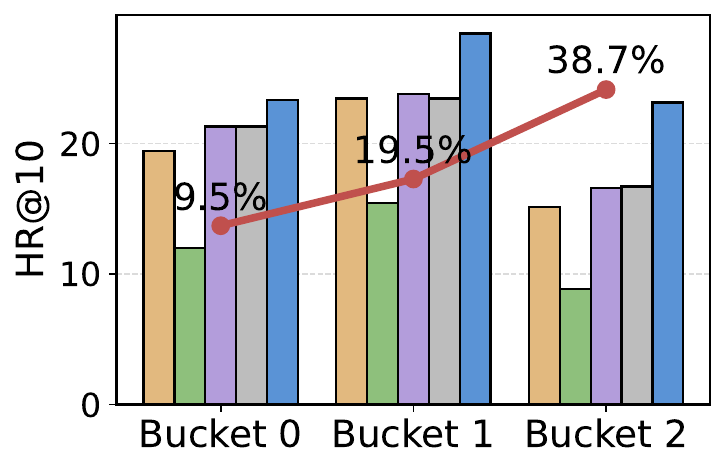}
        \end{subfigure}
        \vspace{-0.5em}
        {\small (a) Domain A}
    \end{minipage}
    \vspace{0.5em}
    
    \begin{minipage}{\columnwidth}
        \centering
        \begin{subfigure}[b]{0.49\columnwidth}
            \centering
            \includegraphics[width=\linewidth]{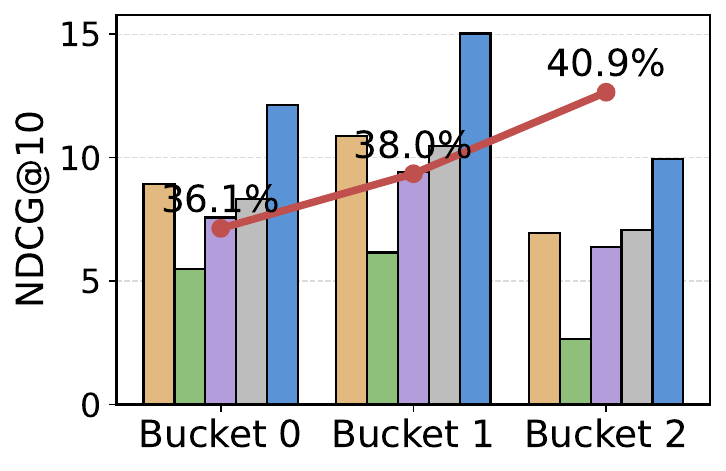}
        \end{subfigure}
        \hfill
        \begin{subfigure}[b]{0.49\columnwidth}
            \centering
            \includegraphics[width=\linewidth]{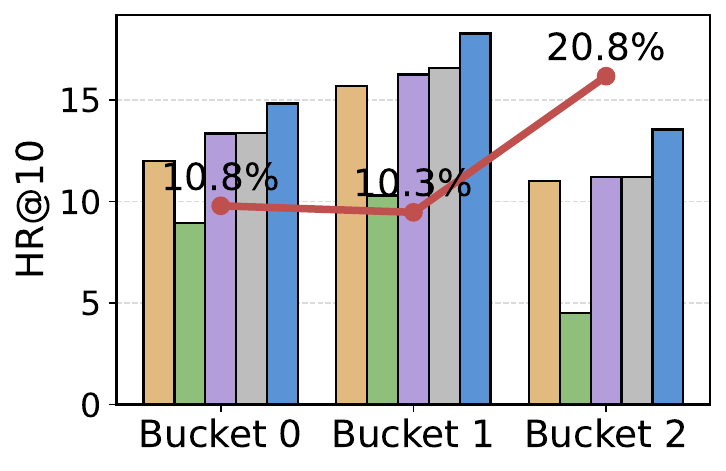}
        \end{subfigure}
        \vspace{-0.5em}
        {\small (b) Domain B}
    \end{minipage}
    \caption{Performance comparison with baselines across different time distributions on the Food-Kitchen dataset.}
    \label{bucket}
\end{figure}

\subsection{Performance across Different Time Distributions (RQ3)}

In this section, we compare BST-CDSR with baseline methods under different temporal distributions. We compute the variance of time intervals for each test sequence to measure interaction uniformity and partition the sequences into three groups, from Bucket 0 (most uniform) to Bucket 2 (most irregular). Since user preferences evolve over time, modeling temporally irregular interactions is inherently more challenging. 

As shown in Figure \ref{bucket}, BST-CDSR outperforms all baseline methods across all domains and evaluation metrics. More importantly, in Bucket 2—the group with the most irregular time distribution—BST-CDSR exhibits the most significant relative advantage over the best baseline method. This is because the baselines do not explicitly model temporal information, limiting their ability to capture interest evolution under irregular interaction patterns. In contrast, BST-CDSR leverages Neural ODEs to model complex temporal patterns, leading to more effective user interest modeling.

It is worth noting that, although sequences with the most uniform temporal distribution (Bucket 0) are theoretically expected to yield the best performance, Bucket 1 achieves the highest results in practice. Further analysis shows that sequences in Bucket 1 have the longest average length, providing sufficient behavioral signals while maintaining a relatively coherent temporal structure. In comparison, Bucket 0 lacks sufficient information due to shorter sequences, while Bucket 2 is harder to model due to severe temporal irregularity.

\subsection{Temporal Semantic Preference Study (RQ4)}

To better understand the contribution of the proposed temporal counterfactual-enhanced module, we conduct a focused study on the semantic branch alone. Specifically, during recommendation we disable all behavioral signals and predict the next item using only the LLM-derived sequence semantic preference. We consider three semantic variants to the LLM: (i) \textbf{Title-only}, which encodes interaction semantics without any temporal information; (ii) \textbf{Title+Time}, which augments each title with discretized time-gap tokens to inject temporal patterns; and (iii) \textbf{+CF-Enhance}, which further applies counterfactual temporal perturbations to explicitly shape the model’s time sensitivity. The results are shown in Table \ref{sematic}.

Compared with \textbf{Title-only}, \textbf{Title+Time} yields slight gains in some domains, yet can also degrade performance in others. In contrast, once we introduce the counterfactual enhancement, performance consistently improves across all domains and all metrics. This observation suggests that once temporal information is introduced, the semantic preferences inferred by the LLM already encode rich temporal characteristics. However, in the absence of explicit constraints, the model may still be biased toward capturing dominant semantic content while underestimating the influence of time on semantics; alternatively, due to the noise in temporal signals, the model can become overly sensitive to incidental temporal variations, leading to the performance fluctuations observed in the \textbf{Title+Time} variant. Conversely, our proposed counterfactual objective effectively constrains temporal semantics, generating more reliable time-aware semantic preferences for recommendations.

\begin{table}[!t]
\centering
\caption{Comparison of different sematic variants on two datasets.}
\vspace{-4mm}
\small
\setlength{\tabcolsep}{3pt}
\renewcommand{\arraystretch}{0.9}

\begin{tabular}{c|ccc|ccc}
\toprule
\toprule
\multirow{2}{*}{\textbf{Sematic Variants}}&\multicolumn{3}{c}{\textbf{Food}}&\multicolumn{3}{c}{\textbf{Kitchen}} \\ 
&{MRR}&{N@10}&{H@10}&{MRR}&{N@10}&{H@10}\\
\midrule
Title only  &0.1638&0.1710&0.2202&0.1053&0.1071&0.1375 \\
Title+Time &0.1649&0.1721&0.2221&0.1040&0.1051&0.1353 \\
+CF-Enhance &\textbf{0.1658}&\textbf{0.1724}&\textbf{0.2259}&\textbf{0.1066}&\textbf{0.1087}&\textbf{0.1393} \\
\toprule
\multirow{2}{*}{\textbf{Sematic Variants}}&\multicolumn{3}{c}{\textbf{Movie}}&\multicolumn{3}{c}{\textbf{Book}} \\ 
&{MRR}&{N@10}&{H@10}&{MRR}&{N@10}&{H@10}\\
\midrule
Title only  &0.0946&0.0978&0.1362&0.0808&0.0820&0.1040 \\
Title+Time &0.0925&0.0958&0.1364&0.0813&0.0826&0.1065 \\
+CF-Enhance &\textbf{0.1041}&\textbf{0.1070}&\textbf{0.1417}&\textbf{0.0821}&\textbf{0.0842}&\textbf{0.1075} \\
\bottomrule
\bottomrule
\end{tabular}
\label{sematic}
\end{table}

\subsection{Visualization of Weights for Long- and Short-Term Behaviors (RQ5)}

\begin{figure}[t!]
    \centering
    \begin{minipage}{\columnwidth}
        \centering
        \begin{subfigure}[b]{0.49\columnwidth}
            \centering
            \includegraphics[width=\linewidth]{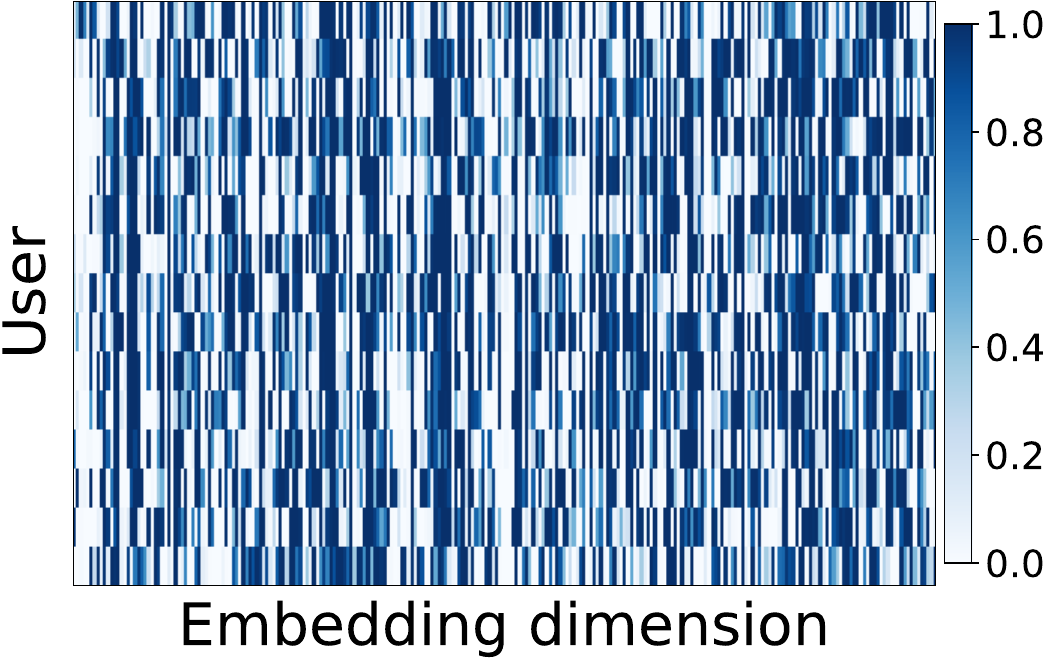}
            {\small (a) Domain A}
        \end{subfigure}
        \hfill
        \begin{subfigure}[b]{0.49\columnwidth}
            \centering
            \includegraphics[width=\linewidth]{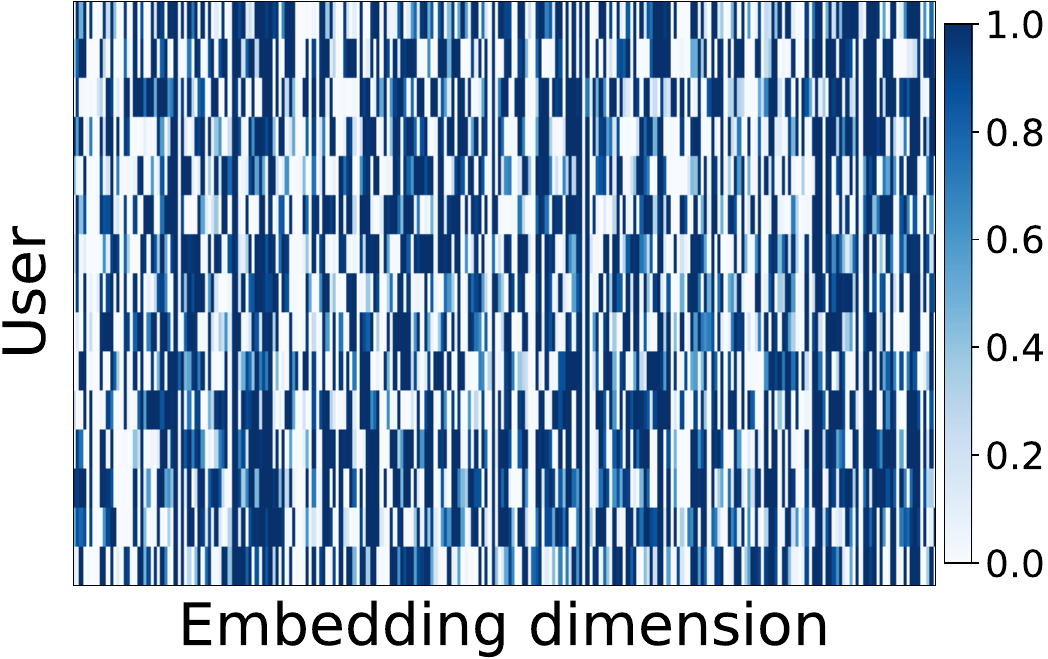}
            {\small (b) Domain B}
        \end{subfigure}
        \vspace{-0.5em}
    \end{minipage}
    \caption{Visualization of weights for combining long- and short-term behavioral preferences on the Food-Kitchen dataset.}
    \label{fuse_weight}
\end{figure}

To further verify whether the proposed behavioral preference evolution module effectively decouples long-term and short-term user preferences, we visualize the learned fusion weight vectors. Figure \ref{fuse_weight} presents the dimension-wise fusion weights for randomly sampled users in both Domain A and Domain B. 

From the visualization, we can observe that the fusion gates exhibit substantial variation across embedding dimensions, indicating that different dimensions selectively emphasize long-term or short-term states, demonstrating dimension-wise decoupling of behavioral preferences. Besides, we observe clear heterogeneity across users. Different users present distinct fusion patterns, suggesting that the model adaptively adjusts the balance between long-term and short-term preferences based on individual behavioral characteristics. Moreover, the overall gate distributions remain stable across both domains, while exhibiting domain-specific patterns. This suggests that the module learns a generalizable mechanism for decoupling long- and short-term preferences, while retaining sufficient flexibility to adapt to different behavioral contexts.

\subsection{Negative Transfer Study (RQ6)}

In this section, to investigate the robustness of BST-CDSR against noise (which typically induces negative transfer), we randomly inject 10\% and 20\% noise interactions into the training set. The results are shown in Table \ref{negative}. As expected, performance degrades with increasing noise, yet it remains superior to the best baseline on most metrics.

This robustness benefits from three aspects. First, the behavioral preference evolution module explicitly decouples long- and short-term preferences under irregular time intervals, reducing the probability of noisy interactions dominating preference representations. Second, incorporating temporal features into semantic preferences explicitly distinguishes semantics driven by situational contexts and user preferences. Finally, the time-preference guided transfer module adaptively assigns cross-domain weights, suppressing unreliable cross-domain information.

\section{RELATED WORK}
\subsection{Cross-Domain Sequential Recommendation.}
Cross-domain recommendation (CDR) leverages auxiliary-domain information to relieve interaction sparsity in the target domain \cite{cdr1, cdr2, cdr3, cdr4}, and cross-domain sequential recommendation (CDSR) further incorporates sequential signals to capture users’ evolving interests \cite{x-cross, autocdsr, gmflowrec, dcdir}. Classic CDSR studies primarily treat completely overlapping users as bridges for cross-domain information transfer \cite{horizonrec, weaverec, aread, da-dan}. As a pioneering work, $\pi$-Net \cite{pai-net} learns sequential user preferences in each domain and transfers cross-domain signals via dedicated sharing units. Based on this, C2DSR \cite{c2dsr} enhances cross-domain consistency by introducing contrastive learning to align user representations. To reduce noisy transfer, RL-ISN \cite{rl-isn} formulates CDSR as a hierarchical reinforcement learning problem for adaptive knowledge selection. Beyond pure sequences, HGTL \cite{hgtl} incorporates category semantics and leverages a cross-domain heterogeneous graph to strengthen structural alignment. More recently, EXHANS \cite{exhans} revisits the optimization process and improves CDSR training via exploration--exploitation driven hard negative sampling. Although existing methods have achieved some advantages, they all simply assume that the intervals across domains are equidistant in time, ignoring domain-specific interaction frequencies and interest decay rates.

\begin{table}[!t]
\centering
\caption{Comparison with the optimal baseline after adding various ratios of noise.}
\vspace{-4mm}
\small
\setlength{\tabcolsep}{3pt}
\renewcommand{\arraystretch}{0.9}

\begin{tabular}{c|ccc|ccc}
\toprule
\toprule
\multirow{2}{*}{\textbf{Models}}&\multicolumn{3}{c}{\textbf{Food}}&\multicolumn{3}{c}{\textbf{Kitchen}} \\ 
&{MRR}&{N@10}&{H@10}&{MRR}&{N@10}&{H@10}\\
\midrule
BST-CDSR  &\textbf{0.1949}&\textbf{0.2028}&\textbf{0.2547}&\textbf{0.1213}&\textbf{0.1246}&\textbf{0.1585} \\
+10\% Noise &0.1822&0.1899&0.2417&0.1119&0.1137&0.1450 \\
+20\% Noise &0.1754&0.1826&0.2335&0.1083&0.1121&0.1466 \\
LLM-EDT &0.1493&0.1705&0.2213&0.1047&0.1084&0.1428 \\
\toprule
\multirow{2}{*}{\textbf{Models}}&\multicolumn{3}{c}{\textbf{Movie}}&\multicolumn{3}{c}{\textbf{Book}} \\ 
&{MRR}&{N@10}&{H@10}&{MRR}&{N@10}&{H@10}\\
\midrule
BST-CDSR  &\textbf{0.1383}&\textbf{0.1436}&\textbf{0.1928}&\textbf{0.1123}&\textbf{0.1147}&\textbf{0.1457} \\
+10\% Noise &0.1255&0.1299&0.1746&0.1038&0.1053&0.1371 \\
+20\% Noise &0.1161&0.1204&0.1641&0.0941&0.0956&0.1274 \\
LLM-EDT &0.1095&0.1190&0.1740&0.0917&0.0953&0.1293 \\
\bottomrule
\bottomrule
\end{tabular}
\label{negative}
\end{table}

\subsection{LLM-based Sequential Recommendation.} 
Large language models (LLMs) have achieved remarkable progress in natural language understanding and generation \cite{nlp1, nlp2, nlp3}, and have recently been introduced into recommender systems as powerful semantic knowledge sources \cite{sigir1, sigir2, sigir3}. For example, CLLMRec \cite{cllmrec} leverages LLMs to perform view augmentation and constructs high-quality contrastive samples to improve sequence representation learning. LLMEmb \cite{llmemb} learns LLM-based semantic embeddings to better handle long-tail items in sequential recommendation. Recent studies further explore integrating LLMs into CDSR to complement purely behaviors information with world knowledge and semantic alignment \cite{ctt}. URLLM \cite{urllm} explores a retrieval-and-generation paradigm, integrating collaborative and structural-semantic information while constraining domain-specific generation. LLMCDSR \cite{llmcdsr} uses LLMs to generate pseudo cross-domain interactions from single-domain sequences. LLM4CDSR \cite{llm4cdsr} extracts unified cross-domain semantic item embeddings from textual attributes. LLM-EDT \cite{llm-edt} proposes LLM-driven cross-domain item augmentation and domain-aware profiling to construct more informative cross-domain supervision. Despite these advances, these LLM-based CDSR methods assume that semantics are static, ignoring the temporal validity of semantic preferences.

\section{CONCLUSION}
In conclusion, we propose a time-aware CDSR framework, BST-CDSR. By introducing a behavioral preference evolution module, it decouples long- and short-term preferences over continuous time. Furthermore, the proposed temporal counterfactual-enhanced semantic preference generator effectively incorporates temporal information into semantic representations. Finally, we design a time-preference-guided domain transfer module to achieve personalized cross-domain transfer. Extensive experiments demonstrate that BST-CDSR consistently outperforms state-of-the-art baselines.



\balance
\bibliographystyle{ACM-Reference-Format}
\bibliography{sample-base}

\end{document}

%% file: baseline.tex
\begin{table*}[!t]
\footnotesize
\setlength{\tabcolsep}{4pt} 
\renewcommand{\arraystretch}{1.0} 
\centering
\caption{Performance of BST-CDSR with all baselines on the three datasets.}
\vspace{-4mm}
\begin{tabular}{cc|cccc|cccc|ccc|c|c}
\toprule
\toprule
\multicolumn{1}{c}{\multirow{2}{*}{\textbf{Datasets}}} & \multicolumn{1}{c|}{\multirow{2}{*}{\textbf{Metrics}}} & \multicolumn{4}{c|}{\textbf{Single-domain Methods}} & \multicolumn{4}{c|}{\textbf{CDSR Methods}} & \multicolumn{3}{c|}{\textbf{LLM-based CDSR Methods}} & \multicolumn{1}{c|}{\multirow{2}{*}{\textbf{BST-CDSR}}} & \multicolumn{1}{c}{\multirow{2}{*}{\textbf{impro.}}} \\
\multicolumn{2}{c|}{} & SASRec & TiSASRec & CL4SRec & IOCRec & $\text{C}^2\text{DSR}$ & EA-GCL & Tri-CDR & ABXI & LLMCDSR & LLM4CDSR & LLM-EDT & & \\

\midrule

\multirow{6}{*}{\textbf{Food}}
& MRR  & 0.0917 & 0.0951 & 0.0949 & 0.1260 & 0.1246 & 0.1119 & 0.1195 & 0.1260 & 0.1392 & 0.1410 & \underline{0.1493} & \textbf{0.1949} & 30.54\% \\
& N@5  & 0.0915 & 0.0943 & 0.0928 & 0.1282 & 0.1250 & 0.0731 & 0.1141 & 0.1277 & 0.1418 & 0.1443 & \underline{0.1537} & \textbf{0.1926} & 25.31\% \\
& N@10 & 0.1063 & 0.1080 & 0.1085 & 0.1375 & 0.1338 & 0.0748 & 0.1311 & 0.1390 & 0.1521 & 0.1558 & \underline{0.1705} & \textbf{0.2028} & 18.94\% \\
& H@1  & 0.0475 & 0.0546 & 0.0513 & 0.0902 & 0.0889 & 0.1055 & 0.0682 & 0.0836 & 0.1140 & 0.1155 & \underline{0.1236} & \textbf{0.1639} & 32.61\% \\
& H@5  & 0.1337 & 0.1319 & 0.1320 & 0.1614 & 0.1570 & 0.1199 & 0.1560 & 0.1673 & 0.1746 & 0.1771 & \underline{0.1808} & \textbf{0.2221} & 22.84\% \\
& H@10 & 0.1800 & 0.1744 & 0.1812 & 0.1901 & 0.1843 & 0.1255 & 0.2087 & 0.2027 & 0.2149 & 0.2176 & \underline{0.2213} & \textbf{0.2547} & 15.09\% \\
\midrule

\multirow{6}{*}{\textbf{Kitchen}}
& MRR  & 0.0588 & 0.0640 & 0.0661 & 0.0973 & 0.0830 & 0.0745 & 0.0713 & 0.0833 & 0.1000 & 0.1019 & \underline{0.1047} & \textbf{0.1213} & 15.85\% \\
& N@5  & 0.0541 & 0.0604 & 0.0614 & 0.0957 & 0.0794 & 0.0481 & 0.0662 & 0.0821 & 0.0988 & 0.1007 & \underline{0.1026} & \textbf{0.1168} & 13.84\% \\
& N@10 & 0.0670 & 0.0715 & 0.0738 & 0.1018 & 0.0880 & 0.0486 & 0.0782 & 0.0908 & 0.1049 & 0.1064 & \underline{0.1084} & \textbf{0.1246} & 14.94\% \\
& H@1  & 0.0269 & 0.0321 & 0.0341 & 0.0770 & 0.0569 & 0.0720 & 0.0320 & 0.0514 & 0.0791 & 0.0813 & \underline{0.0820} & \textbf{0.1005} & 22.56\% \\
& H@5  & 0.0805 & 0.0879 & 0.0877 & 0.1129 & 0.1008 & 0.0779 & 0.0993 & 0.1110 & 0.1147 & 0.1171 & \underline{0.1193} & \textbf{0.1338} & 12.15\% \\
& H@10 & 0.1210 & 0.1225 & 0.1262 & 0.1322 & 0.1276 & 0.0795 & 0.1367 & 0.1382 & 0.1408 & 0.1415 & \underline{0.1428} & \textbf{0.1585} & 10.99\% \\
\midrule
\midrule

\multirow{6}{*}{\textbf{Movie}}
& MRR  & 0.0673 & 0.0744 & 0.0710 & 0.0889 & 0.0719 & 0.0786 & 0.0712 & 0.0923 & 0.1035 & 0.1053 & \underline{0.1095} & \textbf{0.1383} & 26.30\% \\
& N@5  & 0.0612 & 0.0690 & 0.0639 & 0.0863 & 0.0670 & 0.0506 & 0.0623 & 0.0880 & 0.0980 & 0.1000 & \underline{0.1047} & \textbf{0.1334} & 27.41\% \\
& N@10 & 0.0742 & 0.0822 & 0.0769 & 0.0950 & 0.0749 & 0.0516 & 0.0795 & 0.1007 & 0.1124 & 0.1143 & \underline{0.1190} & \textbf{0.1436} & 20.67\% \\
& H@1  & 0.0333 & 0.0387 & 0.0350 & 0.0581 & 0.0490 & 0.0756 & 0.0272 & 0.0563 & 0.0740 & 0.0759 & \underline{0.0820} & \textbf{0.1116} & 36.10\% \\
& H@5  & 0.0891 & 0.0983 & 0.0911 & 0.1130 & 0.0837 & 0.0820 & 0.0959 & 0.1186 & 0.1296 & 0.1315 & \underline{0.1355} & \textbf{0.1575} & 16.24\% \\
& H@10 & 0.1295 & 0.1393 & 0.1314 & 0.1400 & 0.1081 & 0.0851 & 0.1491 & 0.1580 & 0.1678 & 0.1703 & \underline{0.1740} & \textbf{0.1928} & 10.80\% \\
\midrule

\multirow{6}{*}{\textbf{Book}}
& MRR  & 0.0457 & 0.0537 & 0.0561 & 0.0768 & 0.0673 & 0.0592 & 0.0617 & 0.0728 & 0.0853 & 0.0869 & \underline{0.0917} & \textbf{0.1123} & 22.46\% \\
& N@5  & 0.0433 & 0.0513 & 0.0523 & 0.0744 & 0.0649 & 0.0386 & 0.0545 & 0.0685 & 0.0820 & 0.0838 & \underline{0.0876} & \textbf{0.1090} & 24.43\% \\
& N@10 & 0.0499 & 0.0577 & 0.0601 & 0.0784 & 0.0694 & 0.0389 & 0.0646 & 0.0747 & 0.0861 & 0.0881 & \underline{0.0953} & \textbf{0.1147} & 20.36\% \\
& H@1  & 0.0236 & 0.0320 & 0.0339 & 0.0634 & 0.0506 & 0.0564 & 0.0314 & 0.0490 & 0.0702 & 0.0712 & \underline{0.0790} & \textbf{0.0950} & 20.25\% \\
& H@5  & 0.0628 & 0.0694 & 0.0704 & 0.0845 & 0.0778 & 0.0629 & 0.0769 & 0.0864 & 0.0947 & 0.0961 & \underline{0.1028} & \textbf{0.1232} & 19.84\% \\
& H@10 & 0.0834 & 0.0890 & 0.0950 & 0.0970 & 0.0917 & 0.0638 & 0.1081 & 0.1056 & 0.1158 & 0.1173 & \underline{0.1293} & \textbf{0.1457} & 12.68\% \\
\midrule
\midrule

\multirow{6}{*}{\textbf{Art}}
& MRR  & 0.1168 & 0.1240 & 0.1218 & 0.1475 & 0.1504 & 0.1233 & 0.1434 & 0.1515 & 0.1847 & 0.1923 & \underline{0.2010} & \textbf{0.2475} & 23.13\% \\
& N@5  & 0.1104 & 0.1173 & 0.1174 & 0.1434 & 0.1468 & 0.0793 & 0.1047 & 0.1057 & 0.1836 & 0.1911 & \underline{0.1998} & \textbf{0.2455} & 22.87\% \\
& N@10 & 0.1275 & 0.1330 & 0.1329 & 0.1569 & 0.1593 & 0.0803 & 0.1324 & 0.1282 & 0.1958 & 0.2031 & \underline{0.2118} & \textbf{0.2580} & 21.81\% \\
& H@1  & 0.0712 & 0.0833 & 0.0720 & 0.1145 & 0.1123 & 0.1195 & 0.0461 & 0.0489 & 0.1472 & 0.1540 & \underline{0.1632} & \textbf{0.2121} & 29.96\% \\
& H@5  & 0.1477 & 0.1587 & 0.1590 & 0.1716 & 0.1790 & 0.1284 & 0.1489 & 0.1608 & 0.2179 & 0.2265 & \underline{0.2347} & \textbf{0.2816} & 19.98\% \\
& H@10 & 0.2006 & 0.2072 & 0.2068 & 0.2132 & 0.2176 & 0.1317 & 0.2268 & 0.2305 & 0.2703 & 0.2792 & \underline{0.2888} & \textbf{0.3201} & 10.84\% \\
\midrule

\multirow{6}{*}{\textbf{Office}}
& MRR  & 0.1053 & 0.1195 & 0.1129 & 0.1437 & 0.1372 & 0.0945 & 0.1355 & 0.1390 & 0.1674 & 0.1729 & \underline{0.1805} & \textbf{0.2170} & 20.22\% \\
& N@5  & 0.0994 & 0.1133 & 0.1075 & 0.1388 & 0.1335 & 0.0609 & 0.1093 & 0.1124 & 0.1653 & 0.1708 & \underline{0.1792} & \textbf{0.2126} & 18.64\% \\
& N@10 & 0.1203 & 0.1301 & 0.1270 & 0.1568 & 0.1450 & 0.0619 & 0.1324 & 0.1345 & 0.1763 & 0.1819 & \underline{0.1901} & \textbf{0.2267} & 19.25\% \\
& H@1  & 0.0515 & 0.0735 & 0.0592 & 0.1027 & 0.0992 & 0.0914 & 0.0488 & 0.0518 & 0.1287 & 0.1352 & \underline{0.1418} & \textbf{0.1763} & 24.33\% \\
& H@5  & 0.1459 & 0.1509 & 0.1533 & 0.1734 & 0.1654 & 0.0988 & 0.1485 & 0.1524 & 0.1912 & 0.1994 & \underline{0.2081} & \textbf{0.2459} & 18.16\% \\
& H@10 & 0.2112 & 0.2128 & 0.2137 & 0.2297 & 0.2012 & 0.1021 & 0.2211 & 0.2212 & 0.2448 & 0.2521 & \underline{0.2629} & \textbf{0.2905} & 10.50\% \\

\bottomrule
\bottomrule
\end{tabular}
\label{baseline}
\end{table*}